\documentclass{ws-ijmpc}

\begin{document}

\markboth{J. J. Tseng, M. J. Lee \& S. P. Li}
{Heavy-tailed Distributions in Fatal Traffic Accidents: Role of Human Activities}

\catchline{}{}{}{}{}
%%%%%%%%%%%%%%%%%%%%%%%%%%%%%%%%%%%%%%%%%%%%%%%%%%%%%%%%%%%%%%%%%%%%

\title{HEAVY-TAILED DISTRIBUTIONS IN FATAL TRAFFIC ACCIDENTS: ROLE OF HUMAN ACTIVITIES}

\author{JIE-JUN TSENG$^*$ and MING-JER LEE}

\address{Institute of Physics, Academia Sinica, Taipei 115 Taiwan\\
$^*$gen@phys.sinica.edu.tw}

\author{SAI-PING LI}

\address{Institute of Physics, Academia Sinica, Taipei 115 Taiwan \&\\
  Department of Physics, University of Toronto, Toronto, Ontario M5S 1A7, Canada}

\maketitle

%\begin{history}
%\received{Day Month Year}
%\revised{Day Month Year}
%\end{history}

\begin{abstract}
Human activities can play a crucial role in the statistical properties of observables
in many complex systems such as social, technological and economic systems.
We demonstrate this by looking into the heavy-tailed distributions of observables in fatal plane
and car accidents.
Their origin is examined and can be understood as stochastic processes
that are related to human activities.
Simple mathematical models are proposed to illustrate such processes
and compared with empirical results obtained from existing databanks.

\keywords{Heavy-tailed distributions; human activities; stochastic processes; traffic accidents.}
\end{abstract}

\ccode{PACS Nos.: 02.50.-r, 89.75.Da, 89.40.-a}

\section{Introduction}
Many complex systems exhibit heavy-tailed distributions in observables that characterize the systems.
Among them are natural hazards such as earthquakes, landslides, wildfires~\cite{hazards}
or man-made disasters like warfares and global terrorism~\cite{war,terror}, etc.
Over the years, researchers have been asking if the heavy-tailed distributions
observed in complex systems imply something interesting
or whether there is a simple explanation to this kind of phenomena.
If heavy-tailed distributions do appear in some of the observables of the system,
how should one proceed the analysis and determine whether
or not it just results from some stochastic processes.
Until very recently, researchers have suggested that one of the mechanisms
to produce such interesting phenomena is indeed related to human activities~\cite{human1,human2,human3}.
In these works, the researchers identified the heavy-tailed distributions of the observables
as results of human activities.
In particular, they have studied the time interval of occurrence in various cases
such as rating of movies, e-mail communication, etc.
We notice that these studies are concerning the effect of human activities on the temporal behavior
of the above cases.
It would be therefore natural to ask if human activities can also affect observables
other than the temporal statistics of these systems; why and how they are affected.

In this paper, we will investigate the role of human activities in the distribution of observables in complex systems associated with human dynamics.
We will make an attempt by studying the distribution of observables such as fatalities and time intervals of occurrences in fatal traffic accidents
and try to understand the role of human activities in these systems.
We will show that the patterns of human activities do affect temporal observables as well as other kinds of observables in these systems.
Understanding the origin of the heavy-tailed distributions of observables in these systems should give us insight to analyze and understand many other systems, such as those mentioned above.
Two kinds of fatal traffic accidents are considered here, namely, plane and car accidents.
Most people would agree that accidents should be understood as some kind of Poisson stochastic processes and it is hard to imagine how they can have correlations with each other.
In a large scale, a traffic accident occurs in New York should have nothing to do with the occurrence of a traffic accident in Moscow.
However, as we will show below, the distributions of some observables in fatal traffic accidents indeed exhibit interesting phenomena similar to systems which possess heavy-tailed distributions in observables that are commonly believed to be governed by some deep underlying principle.
One would therefore eager to know if there are also deeper reasons for such kind of phenomena in fatal traffic accidents.

In recent years, researchers have looked into the subject of traffic accidents
and studied their temporal properties~\cite{airline,aircar}.
In the case of plane accidents, the authors of Ref.~\refcite{airline} found
that the time lag between commercial airline disasters
and their occurrence frequency could be well described by time-dependent Poisson events.
On the other hand, authors of Ref.~\refcite{aircar} have found
that beyond certain timescales the time dynamics of both plane
and car accidents are not Poissonian but instead long-range correlated.
In the study of accidents, the quantities that one can measure are the fatalities in an accident, the number of accidents or fatalities within a certain period and the time interval between consecutive accidents, etc.
One can therefore proceed to investigate many of these observables aside from the temporal properties and to see if they all display heavy-tailed distributions.
We will here show the empirical result of these quantities from the databanks that we obtain from the web.
Although both plane and car accidents that people studied are classified as traffic accidents, they are indeed very different in nature.
First of all, plane accidents that people have analyzed are global events,
i.e., they keep record of all plane accidents around the world.
However, in the car accidents, due to the limitation of empirical data, people usually study the case within a certain region.
One should therefore argue that the occurrence of air traffic accidents shall not be so much affected by factors
such as human circadian and weekly cycles which affect car accidents.
On the other hand, the number of passengers that an aircraft can take varies from a few to a few hundred while a car usually only takes a few passengers.
Therefore, one would expect that plane and car accidents should have very different behaviors in temporal as well as other properties.
We will show below that the results from empirical data indicate that they are indeed very different.
This paper is organized as follows.
In Section~\ref{sec:data}, we present empirical data of plane and car accidents
which exhibit heavy-tailed distributions in some of the observables studied.
We then give an explanation that can reproduce these results based on human activities
and illustrate this with a simple mathematical model in Section~\ref{sec:model}.
Section~\ref{sec:conclusion} is the conclusion.
%%%

\section{Empirical Data}
\label{sec:data}
%%%
\begin{figure}[ph]
\centerline{\psfig{file=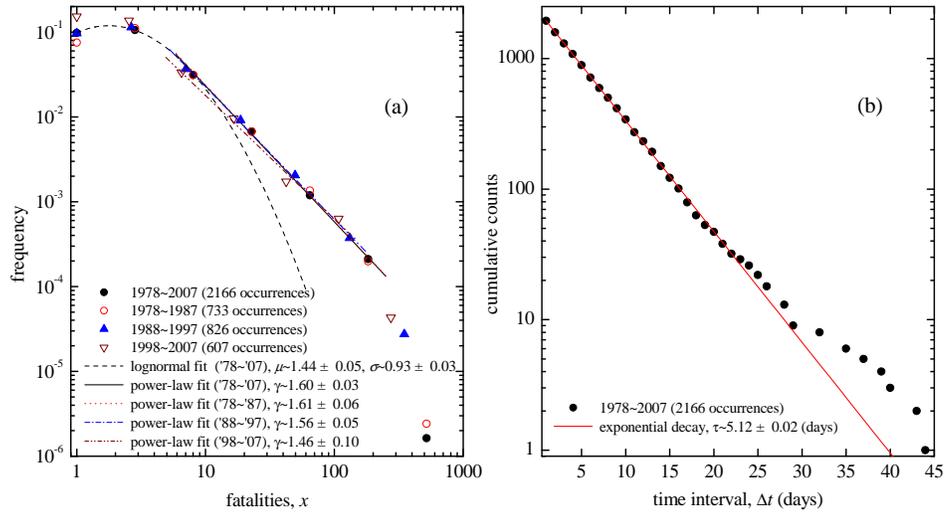,width=0.98\linewidth}}
 \caption{\label{fig:airfatal}(a) A plot of the frequency distribution 
   vs. the fatalities, $x$, in a plane accident during the period 1978-2007.
   The lognormal and power-law fits are also drawn for the comparison.
   (b) Cumulative counts of plane accidents vs. time interval, $\Delta t$,
   between consecutive plane accidents during the same period.
   The exponential fit gives a time constant, $\tau$, equal to 5.12 days.}
\end{figure}
%%%

Fig.~\ref{fig:airfatal}(a) is the frequency distribution of the number of fatalities $x$
involved in plane accidents from 1978 through 2007~\cite{data_plane01,data_plane02}.
One can observe that the empirical distribution deviates from a lognormal one
but rather exhibits a power-law-like ($\sim x^{-\gamma}$) behavior between $x=$ 5 to 200
and drops quickly due to finite size effect (largest possible capacity of current airplanes).
We further divide the plane accidents during this period into 3 different sub-periods,
each lasts for a period of 10 years.
The power-law-like behavior in each of the three curves still holds,
with exponents $\gamma$ ranging between 1.4 and 1.6.
We also note that there is a tendency that the exponents are getting smaller as time goes on.
Fig.~\ref{fig:airfatal}(b) is a semi-log plot of the frequency distribution of time intervals
between consecutive accidents in the same period.
A straight line fit with a time constant, $\tau$, of about 5 days is obtained.
This means that the cumulative counts of plane accidents decay exponentially with respect to the time interval
between consecutive plane accidents, indicating their Poissonian nature.
We further analyze our data by means of the Allen Factor (AF) statistics 
which has been applied to study the temporal behavior of traffic accidents~\cite{aircar}. 
In our analysis, we observe that the AF curve is rather flat up to about five hundred days,
which implies a Poisson-like behavior within such timescales.
The AF fluctuates severely for larger timescales due to the low statistics of our data.
No solid conclusion can therefore be made for the temporal behavior for timescales beyond a few hundred days. 
One should also keep in mind that in these plots, we have ignored incidents from terrorist acts and other incidents,
e.g., those who committed suicides by jumping out of the airplane during the flight, which we do not classify as accidents.
In fact, these incidents only accounted for a small fraction of all incidents and therefore should not affect the result here.
For instance, about 30 incidents related to terrorist attacks were reported in the past ten years (1998-2007)
in a total of 600$+$ incidents, which is about 5\% of all incidents.
The result therefore does not change much by this small fraction.
%%%
\begin{figure}[ph]
\centerline{\psfig{file=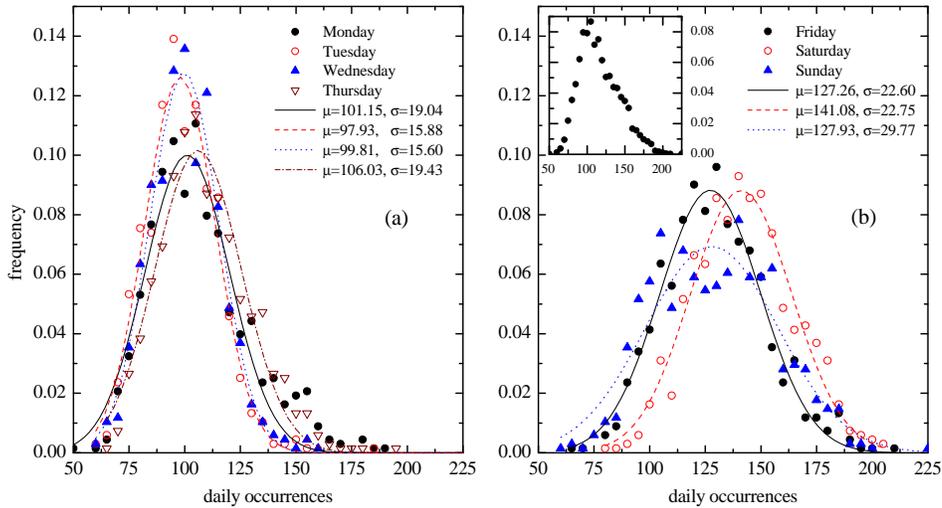,width=0.98\linewidth}}
 \caption{\label{fig:carfatal} Frequency distribution 
   of fatal car accidents in the US for different days of the week
   during the period 1994-2006, (a) Monday-Thursday, (b) Friday-Sunday.
   The inset is a sum of all the distributions in (a) and (b).}
\end{figure}
%%%

The data of fatal car accidents, however, display a totally different scenario.
We here use the dataset of fatal car accidents available on the website of {\it Fatality Analysis Reporting System} (FARS)~\cite{data_car}.
This database contains information about fatal car accidents within the United States in past years and we present here the data from 1994 through 2006 obtained from the website.
The inset in Fig.~\ref{fig:carfatal}(b) is the plot of frequency distribution of fatal car accidents per day from 1994 through 2006.
The shape is somewhat tilted to the left and can be understood as the effect of weekly cycles.
If one plots instead the fatal car accidents that occurred during different days of the week, the distributions can then be very well described by Gaussian distributions.
Fig.~\ref{fig:carfatal}(a) shows plots from Monday through
Thursday and Fig.~\ref{fig:carfatal}(b) shows plots from Friday through Sunday.
The Gaussian fits for different days of the week therefore imply that there should be no correlations among accidents.
Thus, the tilted curve in the inset of Fig.~\ref{fig:carfatal}(b) reflects the pattern of weekly cycles in US citizens¡¦ driving habits (activities) during this period.
In other countries, take Taiwan~\cite{prepare01} for example, the plot of frequency distribution of car accidents per day similar to the inset of Fig.~\ref{fig:carfatal}(b) already gives a good Gaussian fit and the origin can be easily traced back to the driving habits there.

\begin{figure}[ph]
\centerline{\psfig{file=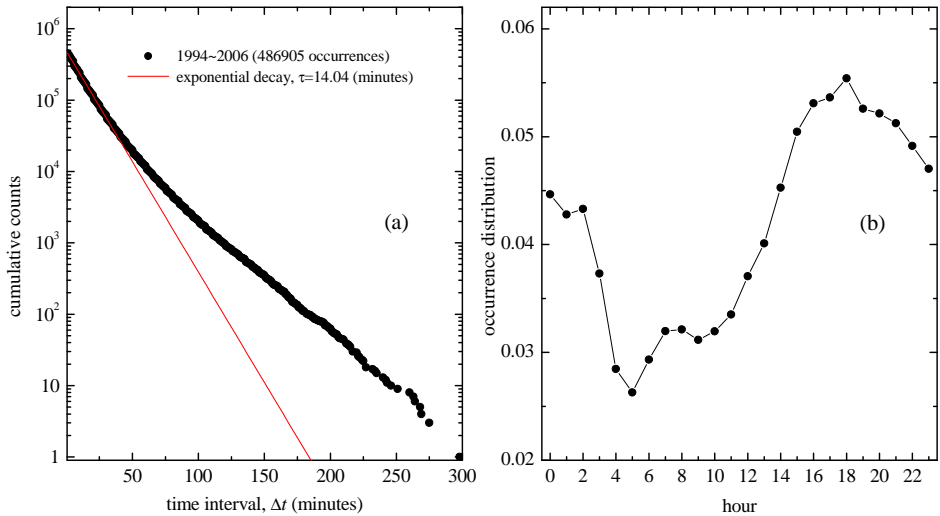,width=0.98\linewidth}}
 \caption{\label{fig:carinterval} (a) A semi-log plot of the frequency
   distribution of the time interval between consecutive fatal car accidents
   in the US during the period 1994-2006.
   (b) The average occurrence distribution of fatal car accidents
   in the US in a 24-hour span during the same period.}
\end{figure}

Fig.~\ref{fig:carinterval}(a) is a plot of the time intervals between consecutive accidents while Fig.~\ref{fig:carinterval}(b) is the average number of fatal car accidents per hour in a 24-hour span from 1994 through 2006.
The distribution of time intervals between consecutive accidents in Fig.~\ref{fig:carinterval}(a) shows a heavy tail significantly different from that of an exponential decay.
This result suggests that there should be other factors involved which render the distribution to differ from that given by a Poisson stochastic process.
%%%
%%%

\section{Model}
\label{sec:model}
In the above, we have observed that in both plane and car accidents, there are heavy-tailed distributions in some observables but exponential decays or Gaussian distributions in other observables that we investigated.
This is however different from usual statistical systems where all the observables will exhibit power-law-like behavior near its critical point.
One would therefore wonder if there are some underlying principles which govern the behavior in the observables of these systems.
We here offer an explanation and illustrate this with a simple model.
We believe that the heavy-tailed distributions in these systems are due to stochastic processes that are related to human activities.
%%%

\subsection{Airplane accidents}
\begin{figure}[ph]
\centerline{\psfig{file=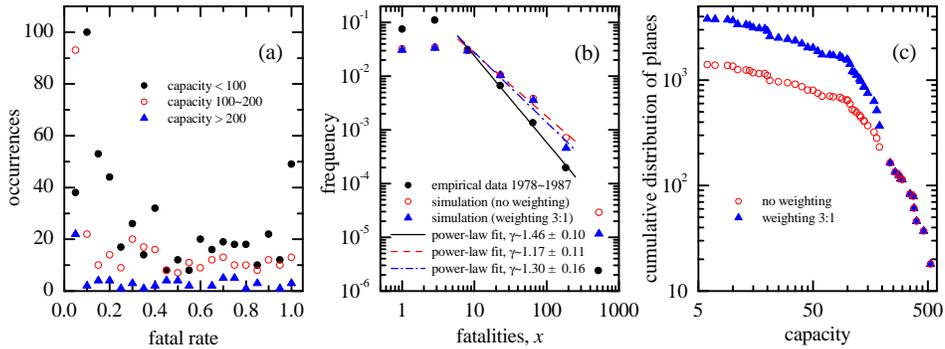,width=0.98\linewidth}}
 \caption{\label{fig:airfatalsim} (a) The distribution of fatality rate.
   The capacities of different makes of airplanes are all normalized to 1
   for comparison.
   (b) shows the simulated result and the empirical data
   during the period 1978-1987,
   while (c) shows the empirical cumulative distribution
   of the capacities of airplanes with different weighting.}
\end{figure}

We begin by postulating that the probability for any plane that flies in the sky to involve in an accident to be the same.
We further assume that the fatalities involved in each accident to be a random number,
i.e., the numbers of fatalities in an accident are uniformly distributed from 1 to $N$, where $N$ is the largest possible capacity of the plane in the accident.
This latter assumption is supported by the empirical data shown in Fig.~\ref{fig:airfatalsim}(a), which are the distributions of fatalities for different makes of airplanes.
The distributions can be reasonably considered as uniform distributions except for the higher frequency of occurrence near zero, which might be due to the fact that some of the planes were indeed used as cargo planes or for training, etc so that there were only a few passengers on the plane.
With these assumptions, we try to simulate the distribution of fatalities involved in plane accidents.
%%%
%%%

Fig.~\ref{fig:airfatalsim}(b) is a plot of the simulated result and the empirical data during the period 1978-1987.
The simulated result gives a power-law behavior with an exponent of about 1.2.
This simulation is carried out by using the empirical distribution of the capacities of airplanes (as shown in Fig.~\ref{fig:airfatalsim}(c)) that were built and started their services before and during the period (1978-87) that we record the plane accidents.
With the two randomness assumptions that we have, the simulated result does reproduce a power-law distribution similar to that of the empirical data.
If one makes further assumptions, the exponent of the simulated result could be improved to a value even closer to the empirical data.
For example, there are many more domestic flights than international flights, which in turn means that the Boeing 727 and 737 planes fly more often than the Boeing 747 planes.
From our first assumption, this implies there are more chances for a Boeing 727 or 737 to get involved in an accident than a Boeing 747.
The dashed line in Fig.~\ref{fig:airfatalsim}(b) is the simulated result with the extra assumption that the ratio of the domestic to international flights to be 3:1.
The exponent now increases to 1.3.
Further assumptions can be made, e.g., by giving a variable between 0 and $N$, the maximum capacity of a plane, which represents the number of passengers that were actually on the plane.
Since our aim is to demonstrate that a power-law-like distribution can result from simple randomness assumptions, we will not pursue further on this.
We should further comment here that as more airplanes were built during the last two decades, the exponent of the power-law behavior of the simulated result gradually gets smaller; a trend that is consistent with the empirical data shown in Fig.~\ref{fig:airfatal}(a).
Since we do not know how the airline companies replace old airplanes with new ones, this is at best a hint that further supports the randomness assumptions.

\subsection{Car accidents}
As shown above, the distributions of the observables studied in car accidents offer a totally different scenario.
While the distribution of fatalities indicates that it is a result of stochastic processes, the time interval plot suggests that there are other possibilities.
We here propose that the heavy-tailed distribution of time interval is indeed due to another pattern of human activities, namely, the human circadian.

\begin{figure}[ph]
\centerline{\psfig{file=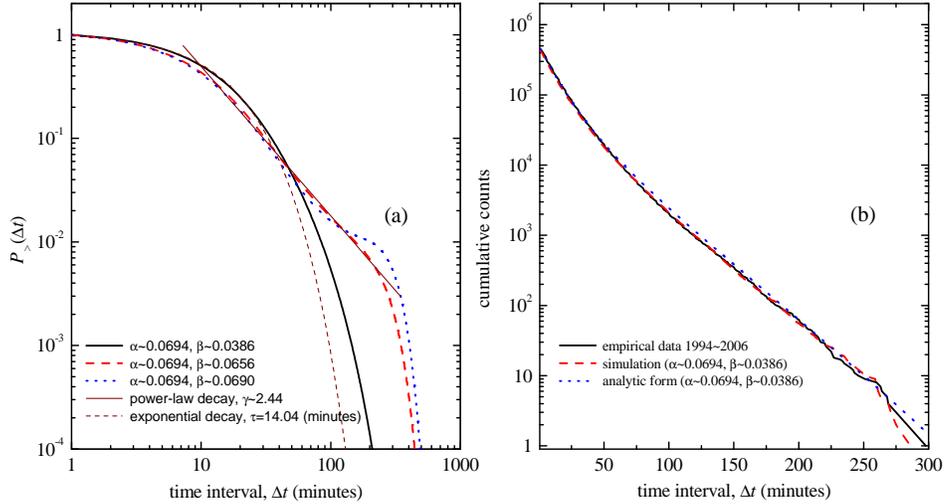,width=0.98\linewidth}}
 \caption{\label{fig:carfatalsim} (a) The plot of $P_{>}(\Delta t)$
  with different $\alpha$ and $\beta$.	
  (b) Fit of the empirical data using $P_{>}(\Delta t)$
  with $\alpha = 0.0694$ and $\beta = 0.0386$.}
\end{figure}

Recall that Fig.~\ref{fig:carinterval}(b) is a plot of the average number of fatal car accidents per hour in a 24-hour span and this value reaches a minimum around 5:00 AM in the morning.
One can imagine that during this time of the day, most American people are still in bed and therefore there are fewer cars on the road than the rest of the day.
We thus propose that the pattern of human circadian or driving habits during a one-day span can be approximated by the following periodic function
%%%
\begin{equation} \label{eqn:period}
  f(t) = \alpha + \beta \sin(2\pi \frac{t}{T})  \,\,\, ,
\end{equation}
%%%
where $t$ is the time variable, $T$ denotes the period
, $\alpha$ is a constant which should be normalized according to the event rate and $\beta$ is the amplitude. 
Although there are many possibilities for the periodic function that can match the curve in Fig.~\ref{fig:carinterval}(b), Eq.(\ref{eqn:period}) is the simplest one which can guarantee that $f(t)$ is always positive with a suitable choice of $\alpha$ and $\beta$.
Eq.(\ref{eqn:period}) can be integrated out analytically and then averaged over, in this case, the 24 hour period.
One can then obtain the cumulative distribution of $\Delta t$ as
%%%
\begin{eqnarray} \label{eqn:intervalfunc}
  P_{>}(\Delta t) &=& \frac{\exp(-\alpha \Delta t)}{2 \pi}
    \int_0^{2\pi} d \theta ~(\alpha+\beta\sin \theta) \nonumber\\
    &\times& \exp\left\{ \frac{\beta T}{2\pi}
      \left[\cos(\theta+ 2\pi\frac{\Delta t}{T} )
        -\cos \theta \right] \right\} \,\,\, ,
\end{eqnarray}
%%%
with $\theta \equiv 2\pi t/T$.
Fig.~\ref{fig:carfatalsim}(a) is a plot of $P_{>}(\Delta t)$ with different $\alpha$ and $\beta$.
Notice that for certain values of $\alpha$ and $\beta$,
Eq.(\ref{eqn:intervalfunc}) could in fact exhibit a heavy-tailed distribution
which is similar to the power-law behavior.
The time intervals of consecutive car accidents in some countries such as Taiwan~\cite{prepare01} indeed display such an interesting power-law behavior.
Fig.~\ref{fig:carfatalsim}(b) is a fit of the empirical data in a semi-log plot using Eq.(\ref{eqn:intervalfunc}) with $\alpha = 0.0694$ and $\beta = 0.0386$.
The dashed line in the same figure is a Monte Carlo simulation of Eq.(\ref{eqn:period}), with the same set of $\alpha$ and $\beta$.
One can see that the empirical data in Fig.~\ref{fig:carinterval}(a) can be well described by the periodic function introduced in Eq.(\ref{eqn:period}).
In addition, we also notice that $P_{>}(\Delta t)$ can develop a bump around the cut-off when $\beta$ becomes large (which means that there exist large fluctuations in the periodic function).
We argue that this bump should be observed in some other complex systems with longer calm-time intervals.
%%%
%%%

\section{Conclusion}
\label{sec:conclusion}
In this paper, we have studied the distributions of the temporal properties
as well as other observables in fatal traffic accidents.
In the case of plane accidents, the time lag of each event
and their number of fatalities could be well described as Poissonian and Gaussian stochastic processes.
On the other hand, the time lag of car accidents
and their occurrence frequency do exhibit heavy-tailed behavior
which is consistent with Ref.~\refcite{aircar}.
We further observe that while the distributions of some observables do suggest that the dynamics of these systems are stochastic, distributions of other observables display instead heavy-tailed behavior.
In our study here, the frequency distribution of the number of fatalities 
$x$ involved in plane accidents exhibits power-law behavior while the frequency distribution of fatal car accidents indicate that these are indeed Poisson stochastic processes.
We here propose an explanation based on stochastic processes that are related to human activities.
Human activities can indeed play a crucial role in the temporal as well as other statistical properties in many complex systems such as social, technological and economic systems.
This is demonstrated explicitly here with the two examples in which the time intervals of occurrences as well as the fatalities are heavy-tailed distributed.
The role of human activities in the temporal properties of complex systems has been pointed out before~\cite{human1,human2,human3}.
We here argue that human activities can affect observables other than the temporal behavior of complex systems.
The distribution of the capacities of airplanes in Fig.~\ref{fig:airfatalsim}(c) is a result of the need of our society which in turn is related to human activities.
In order to understand their effect on observables in a more quantitative manner, we also introduce a periodic function to approximate the pattern of human circadian or driving habits in the study of car accidents.
The empirical data on the temporal properties of car accidents could in fact be well described by this function.
Modifications of this function such as including higher harmonics can be done in a simple manner.
We believe that the heavy-tailed behavior of observables in many complex systems can in fact be understood as stochastic processes that are related to human activities and their temporal properties can be well described by periodic functions of this type.
Work in this direction is in progress~\cite{prepare02}.
%%%

\section*{Acknowledgments}
The research was supported in part by the National Science Council of Taiwan under grants
NSC\#97-2120-M-001-006 and NSC\#97-2112-M-001-008-MY3.
%%%

\end{document}